\documentclass[aps,floatfix,superscriptaddress,twocolumn,showpacs]{revtex4-1}

\usepackage{graphicx}
\usepackage{dcolumn}
\usepackage{bm}
\usepackage{epsfig}

\newcommand\eqn[1]{Eq.\,(\ref{#1})}

\newcommand\fig[1]{Fig.\,{\ref{#1}}}
\newcommand\sect[1]{Sect.\,{\ref{#1}}}

\newcommand\tab[1]{Table~\ref{#1}}

\newcommand{\beq}{\begin{equation}}
\newcommand{\eeq}{\end{equation}}
\newcommand{\bea}{\begin{eqnarray}}
\newcommand{\eea}{\end{eqnarray}}
\newcommand{\hf} {\frac{1}{2}}

\newcommand{\nn}{\nonumber\\}

\def\t{\tilde}

\begin{document}
\title{Critical exponents in quantum Einstein gravity}
\author{S. Nagy}
\affiliation{Department of Theoretical Physics, University of Debrecen,
P.O. Box 5, H-4010 Debrecen, Hungary}
\affiliation{MTA-DE Particle Physics Research Group, P.O.Box 51, H-4001 Debrecen, Hungary}

\author{B. Fazekas}
\affiliation{Institute of Mathematics, University of Debrecen,
P.O. Box 12, H-4010 Debrecen, Hungary}

\author{L. Juh\'asz}
\affiliation{Department of Theoretical Physics, University of Debrecen,
P.O. Box 5, H-4010 Debrecen, Hungary}

\author{K. Sailer}
\affiliation{Department of Theoretical Physics, University of Debrecen,
P.O. Box 5, H-4010 Debrecen, Hungary}

\date{\today}

\begin{abstract}
The quantum Einstein gravity is treated by the functional renormalization group method using
the Einstein-Hilbert action. The ultraviolet non-Gaussian fixed point is determined and its corresponding exponent
of the correlation length is calculated for a wide range of regulators. It is shown that the exponent
provides a minimal sensitivity to the parameters of the regulator which correspond to the
Litim's regulator.
\end{abstract}

\maketitle

\section{Introduction}

The quantum Einstein gravity (QEG) can be a good candidate to describe the gravitational
interactions in the framework of quantum field theory
\cite{Reuter:1996cp,Reuter:2007rv,Reuter:2012id}. The action of the model is integrated
out for all the possible paths of the field variables, which role is now played by the
metrics. The Einstein-Hilbert action of QEG contains only the
cosmological and the Newton couplings \cite{Reuter:2001ag}. The functional renormalization group (RG)
method allows us to perform the path integration systematically for the degrees of freedom, and gives
us the scaling behavior of the couplings starting from the high energy, ultraviolet (UV)
region down to the low energy infrared (IR) regime \cite{Wetterich:1992yh,
Berges:2000ew,Polonyi:2001se}. The method provides an integro-differential
equation for the effective action, the Wetterich equation.
Besides QEG the RG technique is widely used recently in many areas gravitational issues starting
from black holes \cite{Falls:2010he,Koch:2013owa} to cosmological problems \cite{Hindmarsh:2011hx,Kaya:2013bga}.
The QEG is asymptotically safe \cite{Percacci:2007sz,Reuter:2012xf,Nagy:2012ef,Gies:2013pma}, which implies
that there is a UV non-Gaussian fixed point (NGFP) in the phase space with a finite number of
relevant couplings.

The Wetterich equation contains a regulator function to remove the divergences of the momentum
integrals. In the IR limit the effective action should not depend on
the regulator, because it is an artificial term that is put by hand into the action.
However during the solution of the Wetterich equation we use several approximations
and truncations which may introduce some regulator-dependence. In the RG method mostly the optimized
Litim's regulator is used \cite{Litim:2000ci,Litim:2001up}, especially due to its analytic form. The sensitivity of the
physical quantities on the regulator parameters should be minimal \cite{Canet:2002gs}, since the regulator itself
is an artificial element of the RG method and the results should be independent on it ideally.

There are several types of regulators which are widely used, e.g. the power-law, the exponential
and the Litim ones, which can be get by the limiting cases of the so-called compactly supported
smooth (css) regulator \cite{Nandori:2012tc,Nandori:2013nda}. This regulator provides us such a wide class of regulators
which contain all the important types of regulator functions and enables us to investigate
the sensitivity of the physical quantities on a broader range of set of regulators and their parameters.

We calculate the critical exponent $\nu$ of the correlation length and the anomalous dimension $\eta$
around the fixed points of QEG. We obtain that $\eta=2$ independently of the regulators at the UV NGFP, but the value of 
the other exponent $\nu$ can be arbitrary there. The strong truncation of the action might cause such a nonuniversal behavior.
Using the css regulator, we also calculate the value of the exponent $\nu$ for the 3-dimensional $O(1)$ model. 
We investigate how the exponents depend on the css regulator parameters and search for those
parameter regions where the exponents are practically parameter-independent or have at least a minimal
sensitivity of them. We also determine how the exponents scale around the IR fixed point.

The paper is organized as follows. In \sect{sec:model} the investigated models, the RG method,
and the regulators are introduced. In \sect{sec:uv} the UV critical behavior is discussed
for the QEG. The crossover criticality is discussed in \sect{sec:co} for the 3-dimensional $O(1)$
model, while we turn back to QEG  to treat its IR criticality in \sect{sec:ir}.
Finally, in \sect{sec:sum} the conclusions are drawn up.

\section{The model}\label{sec:model}

The Wetterich equation for the effective average action $\Gamma_k$ is \cite{Wetterich:1992yh}
\beq\label{potev}
\dot \Gamma_k = \hf\mbox{Tr}\frac{\partial_t R_k}{\Gamma''_k+R_k}
\eeq
where the dot denotes the derivative w.r.s. to the `RG time' $t=\ln k$,
furthermore the prime is the differentiation with respect to the field
variable and the trace Tr denotes the integration over all momenta and the summation
over the internal indices. The function $R_k$ plays the role of the IR regulator.
The Einstein-Hilbert effective action is
\beq\label{EHtrunc}
\Gamma_k = \frac1{16\pi G_k}\int d^d x\sqrt{g}(-R+2\Lambda_k),
\eeq
with the dimensionful Newton constant $G_k$ and the cosmological constant $\Lambda_k$.
The RG equations are formulated by the dimensionless couplings, i.e.
$\lambda = \Lambda_k k^{-2}$ and $g = G_k k^{d-2}$. Since the determinant of the metric occurs only in Eq. \eqn{EHtrunc}
and nowhere does it any more in our paper, the usage of $g$ for the Newton coupling below shall not be confusing.
The action contains the first two terms in the Taylor expansion of the curvature $R$.
If one inserts \eqn{EHtrunc} into the Wetterich equation then one obtains the
evolution equations for the couplings, which are derived and given in \cite{Reuter:2001ag} and have the form
\bea
\dot\lambda &=& 2(2-\eta)\lambda+\hf(4\pi)^{1-d/2}g\nn
&&\times [2d(d+1)\Phi^1_{d/2}(-2\lambda)-8d\Phi^1_{d/2}(0)]\nn
&&-d(d+1)\eta\t \Phi^1_{d/2}(-2\lambda)],\nn
\dot g &=& (d-2+\eta)g,
\eea
with the anomalous dimension
\beq
\eta = \frac{gB_1(\lambda)}{1-gB_2(\lambda)}.
\eeq
The functions $B_1(\lambda)$ and $B_2(\lambda)$ are
\bea
B_1(\lambda) &=& \frac13(4\pi)^{1-d/2}[d(d+1)\Phi^1_{d/2-1}(-2\lambda)\nn
&&-6d(d-1)\Phi^2_{d/2}(-2\lambda)-4d\Phi^1_{d/2-1}(0)\nn
&&24\Phi^2_{d/2}(0)],\nn
B_2(\lambda) &=& -\frac16(4\pi)^{1-d/2}[d(d+1)\t \Phi^1_{d/2-1}(-2\lambda)\nn
&&-6d(d-1)\t\Phi^2_{d/2}(-2\lambda)],
\eea
with the threshold functions
\bea\label{thrs}
\Phi_n^p(\omega) &=& \frac1{\Gamma(n)}\int_0^\infty dy y^n\frac{r-yr'}{(y(1+r)+\omega)^p},\nn
\t\Phi_n^p(\omega) &=& \frac1{\Gamma(n)}\int_0^\infty dy y^n\frac{r}{(y(1+r)+\omega)^p},
\eea
where $y=p^2/k^2$ and $r=r(y)$ is the dimensionless regulator $r=R/p^2$. 
The dimensionless css regulator has the form
\beq
\label{rcss}
r_{css} = \frac{s_1}{\exp[s_1 y^b/(1 -s_2 y^b)] -1} \theta(1-s_2 y^b),
\eeq
where $b\ge 1$ and $s_1$, $s_2$ are positive parameters.
Although this regulator cannot provide analytic form of the evolution equations,
it gives a broader range of regulators.
The limiting cases of the css regulator provide us the following commonly used regulator functions
\bea\label{limregs}
\lim_{s_1\to 0} r_{css} &=& \left(\frac{1}{y^b} -s_2\right) \theta(1-s_2 y^b),\nn
\lim_{s_1\to 0,s_2\to 0} r_{css} &=& \frac1{y^b},\nn
\lim_{s_2\to 0} r_{css} &=& \frac{s_1}{\exp[s_1 y^b]-1}.
\eea
where the first limit gives the Litim's optimized regulator when $s_2=1$, the second gives the 
power-law regulator, and the third gives the exponential one, if $s_1=1$.
It provides us a possibility of simultaneous optimization
of some physical quantities among the Litim's, the power-law and the exponential
regulators which can be continuously deformed from one to the other by only
two parameters $s_1$ and $s_2$. We note that the $b=1$ case satisfies the normalization
conditions \cite{Reuter:2001ag}
\beq\label{reglim}
\lim_{y\to 0} yr = 1 ~\mbox{and}~\lim_{y\to\infty} yr = 0.
\eeq

\section{Ultraviolet criticality}\label{sec:uv}

\begin{center}
\begin{figure}[ht]
\epsfig{file=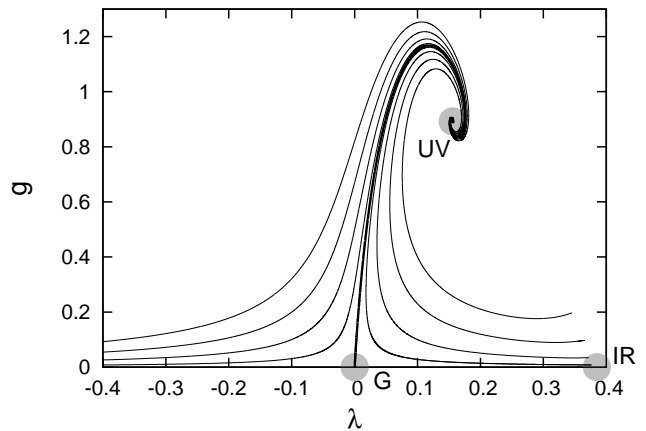,width=6cm,angle=-90}
\caption{\label{fig:qegphase}
The phase structure of QEG is shown for css regulator with the parameters $s_1=s_2=b=1$.
The grey points denote the critical points. The thick line represents the separatrix.
}
\end{figure}
\end{center}

\begin{center}
\begin{figure}[ht]
\epsfig{file=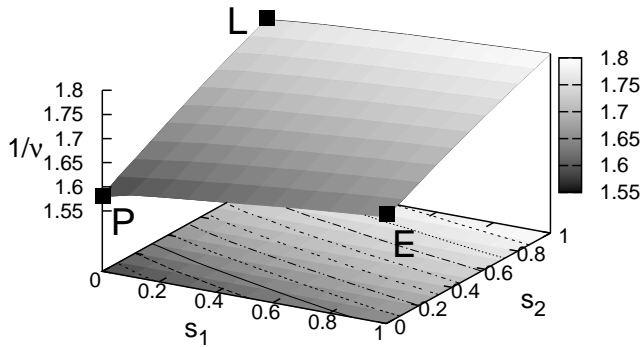,width=6cm,angle=-90}
\caption{\label{fig:sur}
The reciprocal of the exponent $1/\nu$ is calculated for various parameters of
the css regulator, with $b=3$. The value of $1/\nu$ at the corner $s_1=s_2=0$ corresponds to
{\bf P}ower-law regulator result, the corner $s_1=0,~s_2=1$ gives the {\bf L}itim's regulator
result and the corner $s_1=1,~s_2=0$ results in the {\bf E}xponential regulator value.
}
\end{figure}
\end{center}

\begin{center}
\begin{figure}[ht]
\epsfig{file=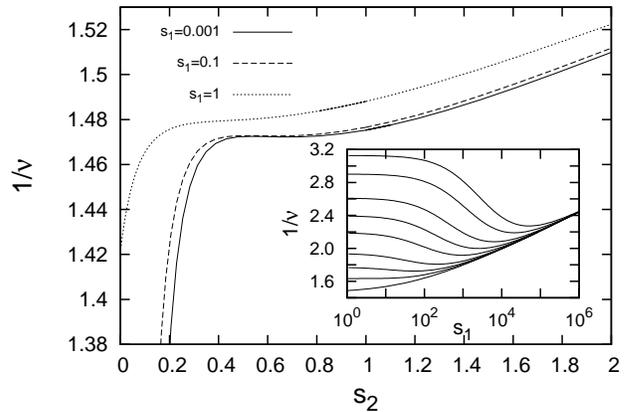,width=6cm,angle=-90}
\caption{\label{fig:qegnulin}
The exponent $1/\nu$ is calculated for various values of parameters of the css regulator.
In the inset the different curves correspond to different values of
$s_2=1,5,10,20,50,100,200,500,1000$, increasing from below. There we chose $b=1$.
}
\end{figure}
\end{center}

QEG has two phases, as is shown in \fig{fig:qegphase}.
The phase space contains 3 fixed points. The UV NGFP is said to be UV attractive,
however it repels the trajectories from the viewpoint of the RG flow. The trajectories start
there and flow towards the hyperbolic Gaussian fixed point (GFP). The trajectories with positive $\lambda$ in the
IR limit constitute the (symmetry-) broken phase of the model which contains an attractive IR fixed point
\cite{Donkin:2012ud,Christiansen:2012rx,Nagy:2012rn,Demmel:2012ub}. The
other trajectories belong to the symmetric phase. One can linearize the RG equations around
the fixed points. In the case of the UV NGFP the eigenvalues of the corresponding stability matrix
can be written as
\beq\label{uveig}
\theta_{1,2}^{UV} = \theta'\pm i\theta'',
\eeq
where the real part of the exponent can be related to the critical exponent $\nu$
of the correlation length, i.e., $\nu=1/\theta'$. We calculated
the position of the fixed point and the corresponding exponent in QEG with Einstein-Hilbert
action in $d=4$. In \fig{fig:sur} we plotted how the reciprocal of the exponent
$1/\nu$ depends on the parameters of the css regulator. The figure demonstrates the appearing
limiting regulators in \eqn{limregs}.
We look for the extremum of $1/\nu$ as the function of $s_1$, $s_2$ and $b$. The increase of
the parameters $s_1$ and $s_2$ gives monotonically increasing exponents, their large
values give logarithmically growing exponents. Therefore, we have to look for a minimum of $1/\nu$.
The limit of the power law regulator $s_1=s_2=0$ gives the minimal value for $1/\nu$, when $b=3$.
The exponent also grows with increasing $b$, therefore one might expect a local extremum at $b=1$.
We calculated $1/\nu$ at the Litim's limit ($s_1\to 0$)
for different values of $s_2$ is shown in \fig{fig:qegnulin} for $b=1$.
Although the optimization of the exponent $\nu$ was performed earlier \cite{Litim:2003vp,Fischer:2006fz},
the figure shows that the css regulator can mimic such a great set of regulators
(including the Litim's, the power-law, and the exponential ones) that it can generate arbitrary values of $1/\nu$
and enables us to perform the optimization program among these fundamental regulators simultaneously.
The UV exponent is known to be $1/\nu=3$ \cite{Hamber:1991ba}, which can be easily reached and exceeded up to practically
$\infty$. It implies that one can tune the regulator parameters to get the physical value of the
exponent, e.g. according to \fig{fig:qegnulin} the choice $b=1$, $s_1=251$ and $s_2=1000$ does the trick.
We note that the strong truncation of the action in the curvature cannot give the real value, therefore
the setting up of regulator parameters in order to get $1/\nu=3$ would not leave us to
the optimal form of the regulator.
Furthermore the exponent shows high sensitivity on the parameters, which also implies that it cannot
be considered as the physical exponent, since we usually look for the parameter regime
where the physical quantities show minimal sensitivity on the regulator \cite{Litim:2001fd,Canet:2002gs}.
We cannot find any absolute extremum in the parameter space of $b$, $s_1$ and $s_2$.
However we have an inflection point at $b=1$, $s_1\to 0$ and $s_2\approx 1/2$, where we have
minimal sensitivity to the regulator parameters.
It corresponds to the Litim's regulator of the form
\beq\label{optreg}
r_{opt} = \left(\frac1{y} -\hf\right) \theta(1-y/2),\nn
\eeq
The value of the exponent there should be considered as the physical exponent, that is
$1/\nu\approx 1.472$, which is less than the half of the real value.
Other RG equations might provide better values for $\nu$ even in the framework
of the Einstein-Hilbert action \cite{Litim:2006dx,Litim:2003vp}, however in this model
if we consider this large set of regulator functions the principle of minimal sensitivity
gives this result. The power-law limit of the css regulator
does not exist for small $b$ in the case of $d=4$, because the RG equation gives
UV divergent integrals, however the finite small value of $s_2$ serves as a UV cutoff of the
loop integral. As $s_2$ gets smaller and smaller the value of $1/\nu$ decreases quickly and
tends to zero and negative values. When $1/\nu=0$ then the eigenvalue $\theta'$ in \eqn{uveig} 
is purely imaginary, then the UV NGFP is not attractive any more and the trajectories evolve
towards a limit cycle as was obtained in \cite{Litim:2012vz,Bonanno:2012dg}. Further decrease of $s_2$ changes
the sign of $\theta'$ to negative making the UV NGFP a UV repulsive one.

We also calculated the position of the fixed point in the phase space. If $s_1\to\infty$
the values of $\lambda^*$ and $g^*$ become $s_2$ independent and scale according to
$\lambda^*\sim s_1^{-0.89}$ and $g^*\sim s_1^{0.89}$ implying that the product $\lambda^*g^*$
tends to a constant, $\lim_{s_1\to\infty}\lambda^*g^* \approx 0.133$. The regulator in \eqn{optreg} gives
$\lambda^*g^* = 0.136$, which value is practically the same. In the limit $s_1\to\infty$
the UV NGFP disappears, since $g^*\to\infty$ making the QEG nonrenormalizable.

\section{Crossover criticality}\label{sec:co}

\begin{center}
\begin{figure}[ht]
\epsfig{file=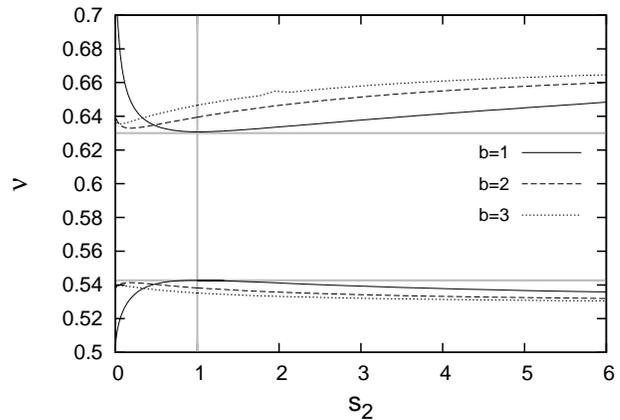,width=6cm,angle=-90}
\caption{\label{fig:d3o1nu}
The exponent $\nu$ is calculated for 2 and 4 couplings and
shown by the curves at the bottom and at the top, respectively. The Litim's regulator gives the
extremal exponent in both cases at the point of the crossing of the grey lines.
}
\end{figure}
\end{center}

There is a crossover (CO) fixed point in QEG, namely the hyperbolic GFP.
However its critical behavior is trivial since the eigenvalues corresponding to the
linearized RG flows equal the negative of the canonical dimension of the
couplings, and the inclusion of further couplings or taking into account
corrections from the gradient expansion beyond the local potential approximation (LPA)
do not change their values. In order to test the CO scaling criticality we treat the 3-dimensional
$O(1)$ model around the Wilson-Fisher (WF) fixed point and the corresponding critical
exponent $\nu$. We start with the effective action
\beq\label{o3d1effac}
\t V = \sum_{i=1}^N \frac{ \t g_i}{(2i)!}\phi^{2i},
\eeq
with the dimensionless couplings $\t g_i$ and restrict ourselves to the truncations $N=2$ and $N=4$.
In LPA the evolution equations for the first two couplings read as
\bea
\dot{\t g}_1 &=& -2 \t g_1 + \t g_2 \bar \Phi^2_{3/2}(\t g_1), \nn
\dot{\t g}_2 &=& -\t g_2 + 6 \t g_2^2 \bar \Phi^3_{3/2}(\t g_1),
\eea
 where the threshold function is introduced according to
\beq\label{thrs2}
\bar \Phi^p_n(\omega) = \frac{1}{(4\pi)^n\Gamma(n)}
\int_0^\infty dy y^{n+1}\frac{r'}{(y(1+r)+\omega)^p}.
\eeq
The evolution equations also contain regulator-dependence. As it is well-known the model plays
the role of the testing ground for any new improvement of renormalization. The exponent $\nu$
of the WF fixed point is well-known, and the optimization of the RG regulator
was first performed around the WF fixed point giving the Litim's regulator, furthermore the value of $\nu$
was calculated for every available type of regulators in the literature.
However by using the css regulator we can get an optimized exponent among
fundamental regulators which was not investigated so far. We considered the RG
evolution for 2 and 4 couplings. The truncation of the scalar potential
to only two terms in the Taylor expansion is very strong, but
we also have only two couplings in QEG with the same level of truncation.
The values of the exponents are plotted in \fig{fig:d3o1nu}.
For 2 couplings we obtained that we have a maximum of the exponent $\nu$ at
$b=s_2=1$ and $s_1\to 0$ which corresponds to the Litim's regulator in \eqn{limregs}.
There the sensitivity of $\nu$ on the regulator parameters is minimal.
Fortunately even the truncation with $N=2$ can provide us a real physical exponent
although it is smaller than the proper value of $\nu$.
We repeated the calculations for $N= 4$ couplings. We got another extremum but now it is
a minimum and it corresponds to the Litim's regulator, too. Further increase of the
number of the couplings and the inclusion of the evolving anomalous dimension $\eta$
may give the Litim's regulator, too.

\section{Infrared criticality}\label{sec:ir}

\begin{center}
\begin{figure}[ht]
\epsfig{file=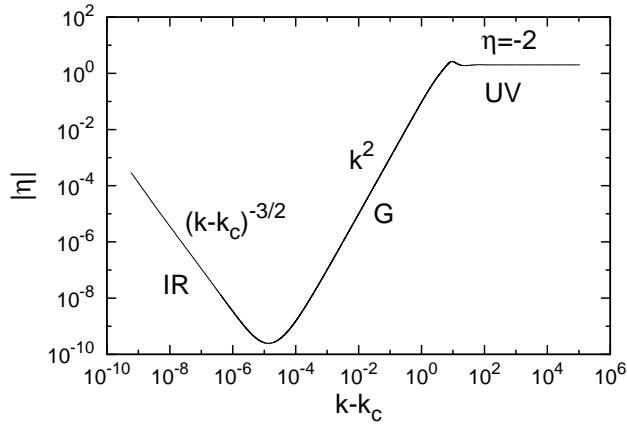,width=6cm,angle=-90}
\caption{\label{fig:qegeta}
The anomalous dimension $\eta$ is plotted as the function of the scale. In the vicinity
of the fixed points the $\eta$ shows different scaling behavior. We chose $s_1=s_2=b=1$
for the css regulator.
}
\end{figure}
\end{center}

In the broken phase of field theoretical models we usually find an IR fixed point.
This is the case in QEG, too \cite{Donkin:2012ud,Christiansen:2012rx,Nagy:2012rn}.
We calculated the exponent $\nu$ by the dynamically induced correlation
\cite{Nagy:2012rn,Nagy:2012np,Nagy:2012qz} around the IR fixed point.
We obtained that the exponent $\nu$ around the IR fixed point equals the one
calculated around the GFP and it is $\nu=1/2$. We also determined the scaling of
the anomalous dimension $\eta$, the results are plotted in \fig{fig:qegeta}.
From the evolution of the Newton constant $g$ it is trivial that in the UV NGFP we have
$\eta=-2$. This corresponds to marginal scaling of $\eta$ in \fig{fig:qegeta}. Then
as the flow approaches the GFP it tends to zero as a power law function $k^2$, where
the exponent 2 is universal.
Going further it starts to diverge as $(k-k_c)^{-\alpha}$, where $\alpha$ is around 3/2 and
shows a slight parameter dependence. The value $k_c$ denotes
the critical scale where the evolution stops due to the appearing singularity.
From the threshold function in \eqn{thrs} in the RG equations the singularity condition is
\beq
y(1+r)-2\lambda = 0,
\eeq
e.g. the $\lambda^*=1/2$, when $r$ is the Litim's regulator. The exponents are summarized
in \tab{tab:exps}.
\begin{table}
\begin{center}
\begin{tabular}{|c||c|c|c|}
\hline
exponent & {\bf UV} & {\bf G} & {\bf IR} \\
\hline
\hline
$\nu$ & 1.472 & 1/2 & 1/2 \\
\hline
$\eta$ & -2 & $k^2$ & $k^{-7/4}$ \\
\hline
\end{tabular}\caption{\label{tab:exps}
The summary of the critical exponents $\nu$ and the anomalous dimensions $\eta$ at the  various
fixed points of QEG.
}
\end{center}
\end{table}
The IR scaling of $\eta$ with the exponent $7/4$ belongs to the Litim's regulator.

\section{Summary}\label{sec:sum}

By using the functional RG method the critical exponent $\nu$
of the correlation length and the anomalous dimension $\eta$ are calculated for the QEG
with the css regulator. We considered the Einstein-Hilbert
action with the Newton constant $g$ and the cosmological constant $\lambda$.
The value of $\nu$ is calculated around the UV NGFP, the GFP, and the IR fixed point.
We obtained that $\nu$ can be arbitrary around the UV NGFP. There can be several reasons
for this result. On the one hand although the regulator is a mass-like term, it possesses a complicated
momentum dependence which might introduce non-local interactions into the action.
Furthermore the deep IR physics cannot be altered by the regulator, but the UV limit can show strong
regulator dependence. On the other hand the Einstein-Hilbert action contains only
two couplings. The inclusion of further couplings in the Taylor expansion in the curvature
\cite{Lauscher:2002sq,Rechenberger:2012pm,Bonanno:2013dja}
may restrict the value of $\nu$ to a certain interval. The several possible extensions
\cite{Eichhorn:2013ug,Henz:2013oxa} can
also give some restriction to the value of $\nu$. We showed that the exponent $\nu=1.472$
can be got if we look for that value which has minimal sensitivity on the regulator
parameters at the Litim's regulator.

The CO scaling around the GFP gives $\nu=1/2$ independently on the approximations
that were used. We demonstrated, that the CO scaling in the 3-dimensional $O(1)$ model
can give a wide range of the value of $\nu$ however an extremum appears in the parameter
space, which corresponds to the Litim's regulator, too.
The IR fixed point gives $\nu=1/2$ which is inherited from the CO GFP.

We also showed that the anomalous dimension takes the value $\eta=-2$ at the UV NGFP, and 
it scales in an irrelevant manner by a universal exponent when the RG flow approaches the GFP,
while it is relevant in the IR, with an exponent which shows moderate dependence on the regulator parameters.

\section*{Acknowledgments}

This research was realized in the frames of T\'AMOP 4.2.4.A/2-11-1-2012-0001 ``National Excellence
Program – Elaborating and operating an inland student and researcher personal support
system'' The project was subsidized by the European Union and co-financed by the
European Social Fund.

\bibliography{nagy}

\end{document}